\documentclass{article}

\usepackage{lineno}
\usepackage{graphicx}
\usepackage{amsmath}
\usepackage[normalem]{ulem}
\usepackage{geometry}
\usepackage{caption}
\usepackage[super,comma,sort&compress]{natbib}
\usepackage[hidelinks]{hyperref}
\usepackage{authblk}
\usepackage{xcolor}
 
\begin{document}


\title{C\text{-}band $160~\text{Gbs$^{-1}$}$ Zero-bias Graphene Photodetectors: Breaking the Responsivity–Bandwidth Trade-off by Heterostructure Engineering}

\author[1]{Karuppasamy Pandian Soundarapandian{*}}

\author[2,3]{Alberto Montanaro\textsuperscript{†}}

\author[6]{Ioannis Vangelidis\textsuperscript{†}}

\author[7]{Stefan M. Koepfli}

\author[1]{Lorenzo Orsini}

\author[1]{Matteo Ceccanti}

\author[7]{Laurenz Kulmer}
\author[2]{Misal Misal}

\author[8]{Tom Reep}

\author[1]{Sebastián Castilla}

\author[9]{Kenji Watanabe}

\author[10]{Takashi Taniguchi}

\author[11]{Seth Ariel Tongay}

\author[8]{Dries Van Thourhout}

\author[7]{Juerg Leuthold}

\author[12,13]{Klaas-Jan Tielrooij}

\author[6,14]{Elefterios Lidorikis}

\author[2,4]{Marco Romagnoli\textsuperscript{‡}}

\author[2,5]{Vito Sorianello\textsuperscript{‡}}

\author[1]{Frank H. L. Koppens{*}}

\affil[1]{ICFO -- Institut de Ci\`encies Fot\`oniques, The Barcelona Institute of Science and Technology, 08860 Castelldefels (Barcelona), Spain}

\affil[2]{CNIT -- National Inter-University Consortium for Telecommunications, 56127 Pisa, Italy}

\affil[3]{TeCIP Institute, Scuola Superiore Sant’Anna, 56124, Pisa, Italy}

\affil[4]{2D Photonics spa, 20900, Monza (MB), Italy}

\affil[5]{Camgraphic srl, 20900, Monza (MB), Italy}

\affil[6]{Department of Materials Science and Engineering, University of Ioannina, Ioannina, Greece}

\affil[7]{ETH Zurich, Institute of Electromagnetic Fields (IEF), Zurich, Switzerland}

\affil[8]{Department of Information Technology (INTEC), Ghent University—IMEC, Ghent, Belgium}

\affil[9]{National Institute for Materials Science (NIMS), Tsukuba, Japan}

\affil[10]{Research Center for Materials Nanoarchitectonics, NIMS, Tsukuba, Japan}

\affil[11]{Arizona State University, Tempe, Arizona, USA}

\affil[12]{Catalan Institute of Nanoscience and Nanotechnology (ICN2), CSIC and BIST, Barcelona, Spain}

\affil[13]{Department of Applied Physics, TU Eindhoven, Eindhoven, Netherlands}

\affil[14]{University Research Center of Ioannina (URCI), Institute of Materials Science and Computing, Ioannina, Greece}

\date{}
\maketitle

\begingroup
\renewcommand{\thefootnote}{\fnsymbol{footnote}}
\footnotetext[2]{These authors contributed equally to this work.}
\footnotetext[3]{the work has been performed while employed at CNIT, PNTLab}
\footnotetext[1]{Corresponding author: karuppasamy.soundarapandian@icfo.eu, frank.koppens@icfo.eu}
\endgroup

\renewcommand\linenumberfont{\normalfont\small}
\begin{abstract}
Graphene photodetectors offer ultrafast response and broadband operation, but their responsivity is typically limited by rapid hot-carrier cooling, leading to a trade-off between sensitivity and speed. Here, we demonstrate that modifying the dielectric environment provides an effective route to control hot-carrier cooling and enhance device performance. By employing a WSe$_2$ encapsulation architecture, we suppress out-of-plane energy dissipation, leading to an increased cooling length ($\sim$2.68 $\mu$m) and a reduced heat-exchange coefficient $\beta$. As a result, we obtain zero-bias graphene photodetectors with responsivities up to $\sim0.12~\mathrm{A/W}$ (Potentially $\sim0.4~\mathrm{A/W}$) while maintaining ultrafast operation beyond setup limited $110~\mathrm{GHz}$. The devices enable direct detection at data rates of $120~\mathrm{Gb\,s^{-1}}$ (NRZ) and $160~\mathrm{Gb\,s^{-1}}$ (PAM-4), with performance achieved using minimal digital signal processing. These results establish dielectric engineering as a key design axis for controlling hot-carrier dynamics, enabling energy-efficient, high-speed optical receivers for next-generation interconnects and AI-driven data systems.

\end{abstract}


\section*{Introduction}

The exponential growth in global data traffic requires low-cost, energy-efficient, high-bandwidth integrated photonic devices capable of sustaining the persistent increase in data rate demands\cite{Cisco2020_AnnualInternetReport,Patrizio2020_800GbE,Cisco2019_400GRollout}. 
With the 1.6~Tbs$^{-1}$ Ethernet standard under development\cite{IEEE8023dj_TaskForce}, photodetectors (PDs) that support simple modulation formats at ultra-high bandwidths are critical for reducing reliance on complex digital signal processing (DSP)\cite{IEEE8023df_TaskForce,Ossieur2025_IntegratedTransceivers,IEEE8023df_B400G_Approval}. The incumbent Ge-on-Si platform achieves responsivities of \(0.3{-}1.0~\mathrm{A/W}\) at bandwidths up to 265~GHz\cite{Lischke2021_UltrafastGePD}, with recent Ge-fin architectures on scalable silicon photonic platforms reaching 240--265~GHz at \(0.3~\mathrm{A/W}\)\cite{Lischke2021_UltrafastGePD}, and co-integrated GeSi-fin photodiodes exceeding 200~GHz in monolithic platforms\cite{steckler2025monolithic}. III--V InP-based modified uni-travelling-carrier (MUTC) photodiodes reach \(0.16{-}0.24~\mathrm{A/W}\) at bandwidths exceeding 220~GHz\cite{Li2024_MUTC,Li2025_MUTC}; however, they require heterogeneous integration with silicon photonics via transfer-printing of III--V coupons, which increases fabrication complexity at scale\cite{Roelkens2024_MicroTransferPrinting}. Critically, both technologies require reverse bias and a transimpedance amplifier (TIA) to convert photocurrent into a usable voltage---a component that dominates receiver power consumption and circuit complexity in massively parallel optical interconnects. In this context, graphene has emerged as a compelling alternative, combining single-layer thickness\cite{Neto2009_ElectronicPropertiesGraphene,Novoselov2004_ElectricFieldEffect}, ultrabroadband absorption\cite{Kuzmenko2008_OpticalConductanceGraphite,Nair2008_TransparencyGraphene,Jang2008_FineStructureGraphene}, high carrier mobility\cite{Bolotin2008_SuspendedGraphene}, and complementary metal-oxide-semiconductor (CMOS) back-end-of-line (BEOL) compatibility\cite{Neumaier2019_IntegratingGraphene} to enable on-chip PDs spanning all telecom bands (O, E, S, C, L, and U)\cite{Agarwal2023_Ultrabroadband,Mak2012_OpticalSpectroscopyGraphene,Soundarapandian2024_SubTHzReceivers}. To date, graphene-based PDs have demonstrated responsivities up to $\sim$0.7~A/W\cite{Ma2020_GrapheneSlotPD,Shiue2015_GrapheneBNPD,Ding2020_UltracompactGraphenePD,Ma2019_PlasmonicEnhancedGraphenePD,Guo2020_GrapheneHybridPD,Liu2021_Si2DPhotodetectors} and bandwidths up to $500\,\text{GHz}$\cite{Ding2020_UltracompactGraphenePD,Ma2019_PlasmonicEnhancedGraphenePD,Koepfli2023_500GHzGraphenePD} by exploiting bolometric\cite{Ma2020_GrapheneSlotPD,Ma2019_PlasmonicEnhancedGraphenePD} and photoconductive\cite{Guo2020_GrapheneHybridPD} effects. These devices, however, operate under applied bias, resulting in substantial dark current and correspondingly high energy consumption\cite{Ma2019_PlasmonicEnhancedGraphenePD,Guo2020_GrapheneHybridPD,Liu2021_Si2DPhotodetectors}.

Conversely, the photothermal effect (PTE) and photovoltaic effect do not require any bias and thus exhibit near-zero dark current. The PTE effect generates a photovoltage from the spatially varying temperature profile\cite{Gabor2014_HotCarrierGraphene}, and PTE-based graphene-integrated PDs (GIPDs) have reached up to $6~\text{V/W}$ responsivity, $110~\text{GHz}$ bandwidth, and 120~\text{Gbs$^{-1}$} data rates\cite{Shiue2015_GrapheneBNPD,Muench2019_WaveguidePlasmonicGraphenePD,Miseikis2020_UltrafastZeroBiasPD,Yu2023_ZeroBiasPNPD,Ma2019_PlasmonicEnhancedGraphenePD,Wang2016_SlotGraphenePD,Schuler2021_MicroringPD,Marconi2021_PAM4GraphenePD}. This performance was driven by efforts to enhance the Seebeck coefficient ($S$) by improving the quality of graphene. On the other hand, plasmonic structures and ring resonators enhance $\nabla$Te through improved absorption\cite{Ioannis_2022} and achieve responsivities up to 12.2 V/W and 90 V/W. However, their bandwidths are limited to 42 GHz and 12 GHz, respectively\cite{Schuler2021_MicroringPD,Schuler2016_PNWaveguideGraphene}. These approaches also present significant drawbacks: (i) plasmonic resonances typically require sub-100~nm lithographic features, making the platform less scalable and highly sensitive to fabrication tolerances, as even minor misalignments can lead to high optical losses or reduced plasmonic enhancement\cite{Alfaraj2025_PlasmonicDevicesReview}; (ii) ring resonators inherently limit the bandwidth of the device, as defined by the optical bandwidth of the ring resonator\cite{Schuler2021_MicroringPD}. Therefore, a trade-off exists between responsivity and bandwidth\cite{Song2011_HotCarrierTransport,Ma2014_HotElectronCoolingGraphene,Yoshioka2022_UltrafastGrapheneDynamics}. Despite extensive work on mobility and absorption, dielectric-environment control of hot-carrier cooling remains largely unexplored in graphene photodetectors. 

In principle, enhancing the $S$ or optical absorption alone does not overcome the intrinsic thermal limitation of the PTE effect. In high-quality graphene (i.e. mobility $> 15{,}000~\text{cm}^2\text{V}^{-1}\text{s}^{-1}$), the $S$ saturates and becomes largely insensitive to further improvements in material quality, making it ineffective for further enhancing responsivity. 
Importantly, this does not imply that higher mobility is detrimental in general; rather, in the PTE regime, increased mobility enhances thermal conductivity, which accelerates hot-carrier cooling and reduces the electron temperature gradient ($\nabla T_e$) driving the photoresponse. As a result, device performance is governed by the generalized heat-exchange coefficient, $\beta = \sqrt{c_e\kappa_e/\tau_c}$, which incorporates both thermal conductivity $\kappa_e$ and the cooling time of hot carriers $\tau_c$. 

Here, we demonstrate that engineering hot-carrier cooling, rather than maximizing carrier mobility, is key to improving zero-bias graphene PDs without compromising bandwidth. We achieve this through an encapsulation-driven strategy that enhances graphene-PD responsivity by engineering $\beta$ via reduced thermal conductivity and prolonged cooling times. Notably, this method does not rely on resonant structures, thereby preserving fabrication simplicity and the intrinsic ultrafast response of graphene. To implement this strategy, we employ tungsten diselenide (WSe$_2$) as both the top and bottom encapsulant (WSe$_2$/Gr/WSe$_2$ - Type~2) in a waveguide-integrated architecture, enabling efficient photovoltage generation via the PTE effect. The effects of WSe$_2$ encapsulation on the electronic and optoelectronic properties of graphene are presented through systematic analysis. We observed that the room-temperature mobility of (WSe$_2$/Gr/hBN) Type~1 heterostructures were higher than the others types reported. We then measured the cooling length ($\xi$) of $\sim2.68~\mu\text{m}$ for the Type~2 heterostructure, almost double that of the Type~1. These contrasting behaviours directly influence photodetector performance since $\beta$ governs the responsivity. Accordingly, we fabricated Type~1 and Type~2 waveguide-integrated PDs, which exhibited responsivities exceeding $0.045~\text{A/W}$ and $0.12~\text{A/W}$, respectively, with no observable roll-off in the frequency response up to $110~\text{GHz}$ at $1550~\text{nm}$ (measurement-limited), consistent with the proposed mechanism. By combining high responsivity and bandwidth, these PDs achieved setup-limited direct detection of non-return-to-zero (NRZ) and PAM-4 optical signals at data rates of $120$ and $160~\text{Gbs$^{-1}$}$, respectively. These findings, together with the developed theoretical model, refine the conventional expectation that higher material quality alone enhances responsivity, by highlighting the competing role of hot-carrier cooling.

\section*{Results}
\subsection*{Transport characterisation}

\begin{figure}[h!]
    \centering
    \includegraphics[width=1\linewidth]{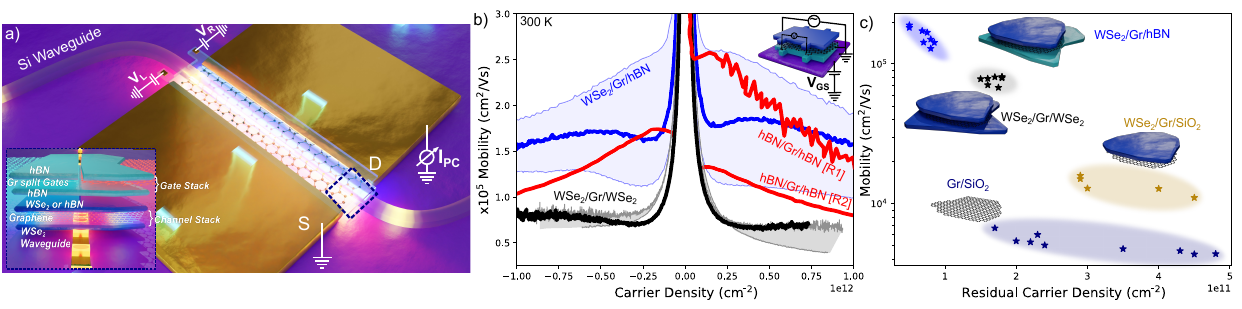}
    \caption{ Heterostructure-dependent electrical quality in graphene. \textbf{a} Schematic of the waveguide-integrated graphene photodetector, with inset showing the van der Waals heterostructure stack. \textbf{b} Carrier mobility for graphene encapsulated Hall-bars with different encapsulants, demonstrating significant variation depending on the heterostructure. The shaded region represents the device-to-device variation range between the minimum and maximum measured value. \textbf{c} Residual charge density ($n^*$) as a function of mobility, establishing the relationship between the mobility and $n^*$ for different heterostructures}
    \label{fig:HallMob}
\end{figure}

Fig. \ref{fig:HallMob}a presents a schematic of the waveguide-integrated graphene photodetector used in this work, with the inset highlighting the heterostructures employed, in particular the channel stack. This structural detail is critical, as the responsivity is governed by $\beta$, which captures the interplay between thermal conductivity and cooling dynamics\cite{Ioannis_2022}. Because the thermal conductivity is directly proportional to carrier mobility, the choice of encapsulating materials in the graphene heterostructure plays a central role in device performance. State-of-the-art graphene encapsulation with hBN is the most common and experimentally accessible strategy for enhancing carrier mobility to values exceeding $150{,}000~\text{cm}^2\text{V}^{-1}\text{s}^{-1}$ \cite{purdie_cleaning_2018, wang_clean_2023,huang_versatile_2020}, thereby enabling $S$ as high as $\sim300~\mu\text{V/K}$\cite{Schuler2021_MicroringPD}. However, the $S$ saturates for mobilities beyond $\sim10{,}000~\text{cm}^2\text{V}^{-1}\text{s}^{-1}$, such that further increases in mobility predominantly enhance thermal conductivity and, consequently, reduce responsivity\cite{Ioannis_2022,Muench2019_WaveguidePlasmonicGraphenePD}. Concurrently, the residual carrier density ($n^*$) becomes a limiting factor. An optimal regime combines moderately high mobility with low $n^*$, providing favourable conditions for efficient photodetection. Mobility and disorder must be balanced rather than maximizing mobility alone. Therefore, to address the aforementioned limitations, we propose WSe$_2$ as an alternative, an atomically flat and potentially scalable 2D material that exhibits weaker hot-carrier-phonon coupling than hBN\cite{Pogna2021_HotCarrierCoolingHighQuality}.

First, we examined the influence of WSe$_2$ encapsulation on graphene mobility. To do so, we fabricated heterostructures with three distinct configurations: WSe$_2$/Gr/hBN (Type 1), WSe$_2$/Gr/WSe$_2$ (Type 2), and WSe$_2$/Gr/SiO$_2$. Subsequently, these heterostructures were fabricated into Hall-bar geometries for four-terminal transport measurements, as shown in the inset of Fig. \ref{fig:HallMob}b. Fig. \ref{fig:HallMob}b shows the measured Drude mobility $\left(\mu = \sigma/(ne)\right)$ as a function of the charge carrier density ($n$) induced by the gate voltage, where Types 1 and 2 are compared with the state-of-the-art reference hBN/Gr/hBN heterostructures R1\cite{huang_versatile_2020} and R2\cite{purdie_cleaning_2018}. Type 1 devices exhibited an average mobility of $\sim155{,}000~\text{cm}^2\text{V}^{-1}\text{s}^{-1}$ for both the electron and hole transport at $n \approx 1 \times 10^{12}~\text{cm}^{-2}$ at room temperature. This exceeds the performance of R1 and R2, highlighting the effectiveness of WSe$_2$ as a capping layer. In contrast, Type 2 devices exhibited a lower value of $\sim80{,}000~\text{cm}^2\text{V}^{-1}\text{s}^{-1}$ at $n \approx 1 \times 10^{12}~\text{cm}^{-2}$. The blue and grey shaded regions in the plot indicate the mobility ranges measured across different devices of Types 1 and 2 heterostructures.

Next, in Fig. \ref{fig:HallMob}c, we compared the $n^*$ values of Types 1, 2, and 3 with those of a standard Gr/SiO$_2$ device to evaluate the disorder introduced by WSe$_2$ encapsulation. The mobility values presented here were measured at $n = 1 \times 10^{12}~\text{cm}^{-2}$. The $n^*$ value was estimated by identifying the intersection point between the linear and flat regions of the $n$-dependent conductivity curve. As anticipated, the Gr/SiO$_2$ device exhibited a high $n^*$ in the range of approximately $1.5$-$5 \times 10^{11}~\text{cm}^{-2}$. However, capping graphene with WSe$_2$ (Type~3) resulted in a comparable $n^*$ range while achieving twice the mobility. Type~2 devices exhibited a more uniform $n^*$ in the narrow range of $1.5$-$1.8 \times 10^{11}~\text{cm}^{-2}$, achieving an approximately fivefold increase in mobility compared to Type~3 devices, clearly highlighting the influence of WSe$_2$ as a substrate. Nevertheless, Type~1 devices have the lowest $n^*$ range of $\sim5$-$9 \times 10^{10}~\text{cm}^{-2}$, indicating minimal induced disorder, leading to the highest mobility among all combinations. The observed relationship between the mobility and $n^*$ aligns with the literature\cite{couto_random_2014,karuppasamy_hysteresis-free_2023} and explains the mobility variations depicted for Types~1 and 2 in Fig. \ref{fig:HallMob}b. As an atomically flat substrate, the $n^*$ of Type~2 devices is expected to closely resemble that of Type~1 devices. The strain induced by the lattice mismatch ($\sim 25\%$)\cite{kaloni_quantum_2014} may explain the observed differences; however, further studies are required, which lie beyond the scope of the present work.

\subsection*{Cooling length determination}
Next, we focused on the thermal management of the hot carriers in the heterostructures discussed above, since \( \Delta T_e \propto \frac{\sinh\!\left(\tfrac{W}{2\xi}\right)}{2W\cosh\!\left(\tfrac{W}{2\xi}\right)} \), with particular emphasis on the cooling length \( \xi = \sqrt{\frac{\kappa_e}{\gamma c_e}} \), where $W$ is the source-drain length, $\kappa_e$ is the thermal conductivity of graphene, $\gamma$ is the electron-lattice cooling rate, and $c_e$ is the specific heat capacity\cite{Song2011_HotCarrierTransport}. Among the various hot-carrier cooling mechanisms, out-of-plane cooling through a substrate is the dominant process in high-quality graphene-based heterostructures (hBN/Gr/hBN). In particular, hyperbolic phonons in hBN-encapsulated devices play a crucial role in enabling efficient cooling\cite{Tielrooij2018_OutOfPlaneHeat}. To measure the $\xi$ for the Type 1 and Type 2 heterostructures, we first fabricated long-channel PDs (See Supp. Note 2) on top of the hBN/Gr bottom split gates and compared the results with those of an hBN/Gr/hBN heterostructure. The electrical and optoelectronic properties of the PDs were determined, and their mobilities, $n^*$, and operating gate voltage (OGV) were carefully extracted (See Supp. Note 3). These long-channel scattering-type scanning near-field optical microscopy (sSNOM) PDs are distinct from the Hall bars (Fig. \ref{fig:HallMob}) and the waveguide-integrated PDs (Fig. \ref{fig:pd}); the open-access channel geometry required by sSNOM probing is incompatible with the waveguide-integrated geometry (see Methods).

\begin{figure}[h!]
    \centering
    \includegraphics[width=1\linewidth]{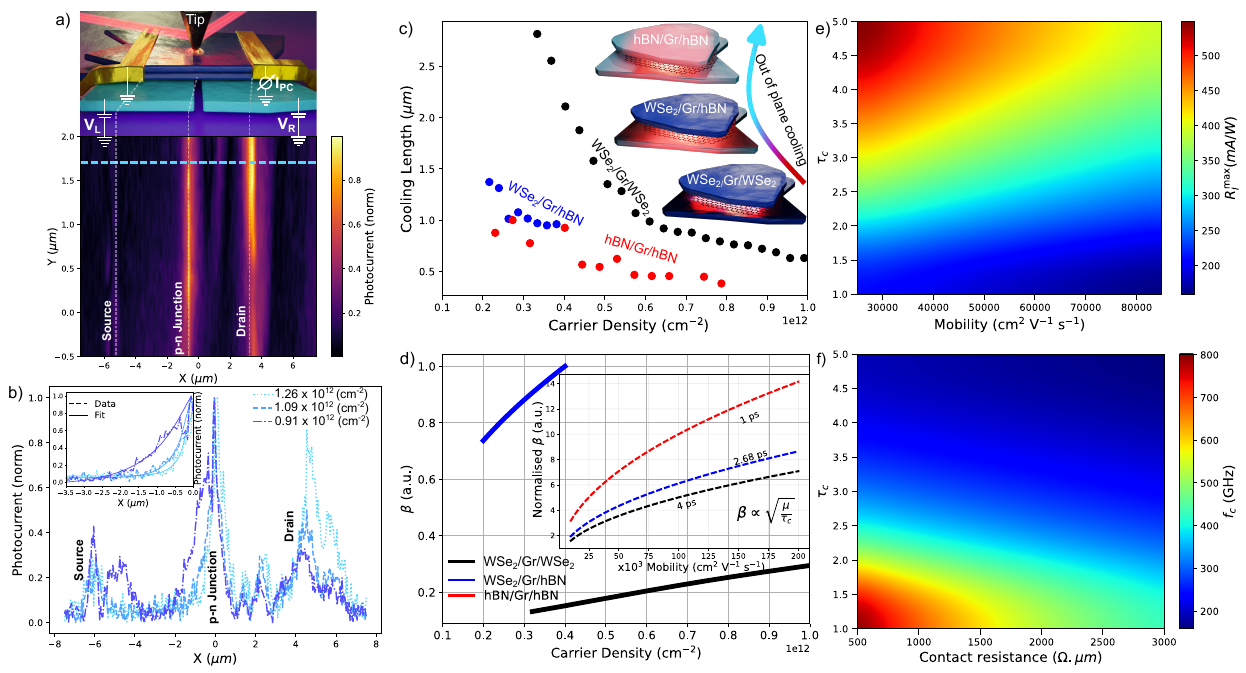}
    \caption{ Cooling length characterisation in high-quality heterostructures. \textbf{a} Schematic of the sSNOM technique (top) and corresponding two-dimensional photocurrent map of the Type~2 heterostructure (bottom). The dashed lines indicate the linecuts used to extract the photocurrent profiles shown in Fig. \ref{fig:cool}b for different carrier densities ($n$) \textbf{b} Line profiles of the photocurrent for three carrier densities ($n$), extracted along the blue line in Fig. \ref{fig:cool}a; the inset shows a representative fit. \textbf{c} Extracted cooling length ($\xi$) as a function of $n$ for three heterostructures (hBN/Gr/hBN, Type 1, Type 2); Type~2 exhibits a longer $\xi$, indicating suppressed hot-carrier cooling. Data points with carrier densities near the residual charge density ($n^*$) are excluded to avoid misleading interpretation near the charge inhomogeneity regime. The inset shows a schematic representation of out-of-plane cooling in the heterostructures. \textbf{d} Normalized heat exchange coefficient $\beta$ calculated from the measured cooling length and the thermal conductivity of Type~1 (blue) and Type~2 (black) heterostructures, plotted against $n$. The inset shows the calculated $\beta$ as a function of mobility for three cooling times 1~ps, 2.68~ps, and 4~ps corresponding to hBN/Gr/hBN, Type~1, and Type~2 heterostructures, respectively. \textbf{e} Analytical upper-bound responsivity $R_I^{\infty} = \frac{g S}{\beta}$ as a function of mobility $\mu$ and cooling time $\tau_c$, evaluated using the closed-form framework (see Methods). \textbf{f} Simulated Bandwidth estimator $f_c = \left(f_{c1}^{-1} + f_{c2}^{-1}\right)^{-1}$ as a function of contact resistance and cooling time, with $C = 2~\text{fF}$ (see Methods).}
    \label{fig:cool}
\end{figure}

The sSNOM tip serves as a near-field point source for localized illumination of the channel. Under illumination from a focused IR laser beam, the sSNOM tip exhibits the enhanced light-induced phenomenon known as the “lightning rod effect.” This effect generates a highly localized hotspot of light at the apex of the scanning tip\cite{Pohl1991_SNOM,woessner_highly_2015}. This technique enables spatially resolved photoresponse at the nanoscale. Fig. \ref{fig:cool}a (top) shows a schematic representation and measurement configuration of the fabricated devices. Gate voltages were applied to the bottom split gates to create a p-n junction. Spatial scans of the PDs were performed while measuring the resulting photocurrent, as shown in Fig. \ref{fig:cool}a (bottom). The map revealed three distinct regions of photoresponse corresponding to the graphene/metal interface and the p-n junction formed. This process identified a clean region for extracting $\xi$, as depicted by the dotted blue line in the photocurrent map. Next, we performed a gate-voltage-dependent line scan along the blue dotted line while simultaneously sweeping the split-gate voltage simultaneously, that is (left gate = right gate), the p-n and n-p regions of the PTE response, where the photocurrent is maximal. Fig. \ref{fig:cool}b presents the line cuts from the Type~2 device, acquired at $n$ = 0.91, 1.09, and 1.25$ \times 10^{12}~\text{cm}^{-2}$, respectively. Increasing gate voltage enhances exponential decay near the p–n junction. This indicates a reduced cooling length $\xi$. The inset shows the exponential fit (solid line) obtained to extract the $\xi$ from the gate-voltage-dependent line scans (dotted lines).

Fig. \ref{fig:cool}c shows the $\xi$ of the Type~1, Type~2, and hBN/Gr/hBN PDs as a function of the $n$. The $\xi$, extracted from the gate-voltage-dependent line-scan maps (See Supp. Note 3), was $\sim1~\mu\text{m}$, $\sim1.4~\mu\text{m}$, and $\sim2.68~\mu\text{m}$ for the hBN/Gr/hBN PDs, Type~1, and Type~2, respectively. Values of $\xi$ below $n^*$ were disregarded to ensure fair and practical consideration of the operational $n$ of a PD. At higher carrier density, hot-carrier cooling and optical phonon cooling becomes more efficient, leading to shorter $\xi$\cite{Pogna2021_HotCarrierCoolingHighQuality,Tielrooij2018_OutOfPlaneHeat}. hBN/Gr/hBN PDs exhibit the lowest $\xi$ at even small $n$; this is due to the stronger out-of-plane cooling of the hot carriers into hBN, aided by hyperbolic phonons\cite{Tielrooij2018_OutOfPlaneHeat}. By contrast, Types~1 and 2 exhibit longer $\xi$, reflecting reduced heat-dissipation pathways (substrate phonon coupling) due to partial or complete removal of hBN encapsulation. The systematic increase in $\xi$ observed across the heterostructures provides strong experimental evidence that dielectric environment governs hot-carrier cooling in graphene, consistent with previous studies\cite{Tielrooij2018_OutOfPlaneHeat,Pogna2021_HotCarrierCoolingHighQuality,wang_dielectric_2026}. Type~2 exhibits the longest cooling length among the reported graphene heterostructures\cite{Tielrooij2018_OutOfPlaneHeat,Betzac_hot_elec_2012,Gabor2014_HotCarrierGraphene,graham_photocurrent_2013}. Suppressed coupling to substrate phonons shifts hot-carrier relaxation toward optical-phonon-mediated cooling, a process sustained by carrier–carrier scattering through rapid carrier re-thermalization\cite{Pogna2021_HotCarrierCoolingHighQuality}. In the Type~2 heterostructure, the dielectric environment is expected to weaken carrier–carrier scattering, thereby reducing the efficiency of optical phonon emission and suppressing hot-carrier cooling as seen in recent studies\cite{wang2026dielectric}$^{\mathrm{,Yishu\ et\ al.,\ arXiv,\ 2026}}$. 

The extracted cooling length outside the hydrodynamic heat-flow-dominated Dirac fluid regime\cite{block_observation_2021}, together with the measured conductivity of the heterostructures, was used to determine the cooling time ($\tau_c$), yielding average values of $\sim$1 ps, $\sim$2.6 ps and $\sim$4 ps for hBN/Gr/hBN, Type~1 and Type~2 around the OGV (See Supp. Note 1), respectively, consistent with the literature\cite{Tielrooij2018_OutOfPlaneHeat,Pogna2021_HotCarrierCoolingHighQuality}. 
To make the role of thermal transport explicit, $\beta$ is estimated from the experimentally extracted $\xi$ and the thermal conductivity inferred from $\sigma$, providing a physically consistent metric for comparing hot-carrier cooling across heterostructures (Fig. \ref{fig:cool}d). The Type 2 heterostructure exhibits a lower $\beta$ across the operating range compared to Type 1. The inset illustrates the scaling of $\beta$ with $\mu$ and $\tau_c$, showing that the reduced $\beta$ in Type 2 arises from its extended cooling dynamics despite lower mobility. Substituting the measured $(\mu, \tau_c, n^*)$ of each stack into the contact-dominated scaling $R_I^{\infty} \propto \sqrt{\tau_c/\mu}$ established in the Introduction (with the contact-dominated regime $R_C \gg R_G$ verified experimentally in the DC characterization below), Type~2's longer $\tau_c$ outweighs its lower mobility, exactly the trade-off Type~2 wins over Type~1. Using the measured $\mu$ and extracted $\tau_c$, we evaluated the analytical PD responsivity (See Supp. Note 1), thereby probing the upper limit of the responsivity in the engineered heterostructures as shown in Fig. \ref{fig:cool}e. Given the enhanced \( \tau_c \) observed in these heterostructures and the dominance of contact resistance (R$_\text{c}$) over channel resistance in our devices, we evaluated the analytical frequency response (f$_c$) as a function of \( \tau_c \) and R$_\text{c}$ to estimate the achievable bandwidth. Fig.~\ref{fig:cool}f was obtained by considering $f_c = \left(f_{c1}^{-1} + f_{c2}^{-1}\right)^{-1},\; f_{c1} = \frac{1}{2\pi R_{D}C}\,\; f_{c2} = \frac{1}{\tau_c}$, with $R_D = R_G + R_C$ the total device resistance and $C = 6~\mathrm{fF}$, consistent with previously measured devices and the literature\cite{Marconi2021_PAM4GraphenePD,Ioannis_2022}. This, in turn, enables rational selection of channel length and width for the PD, subject to the constraint that the PD length should be $<\xi$\cite{ma_competing_2014}. Simulations predict responsivities above 0.4~A/W at 250-300 GHz for Type~2 devices. Type~1 devices reach 0.3-0.4 A/W at 350–450~GHz. Thus, the systematic reduction of $\beta$ across heterostructures, together with the concurrent increase in cooling length, provides direct experimental evidence that the dielectric environment, not carrier mobility alone, governs hot-carrier cooling, and thereby the resulting responsivity.


\subsection*{DC characterisation of the waveguide-integrated PD}

\begin{figure}[h!]
    \centering
    \includegraphics[width=1\linewidth]{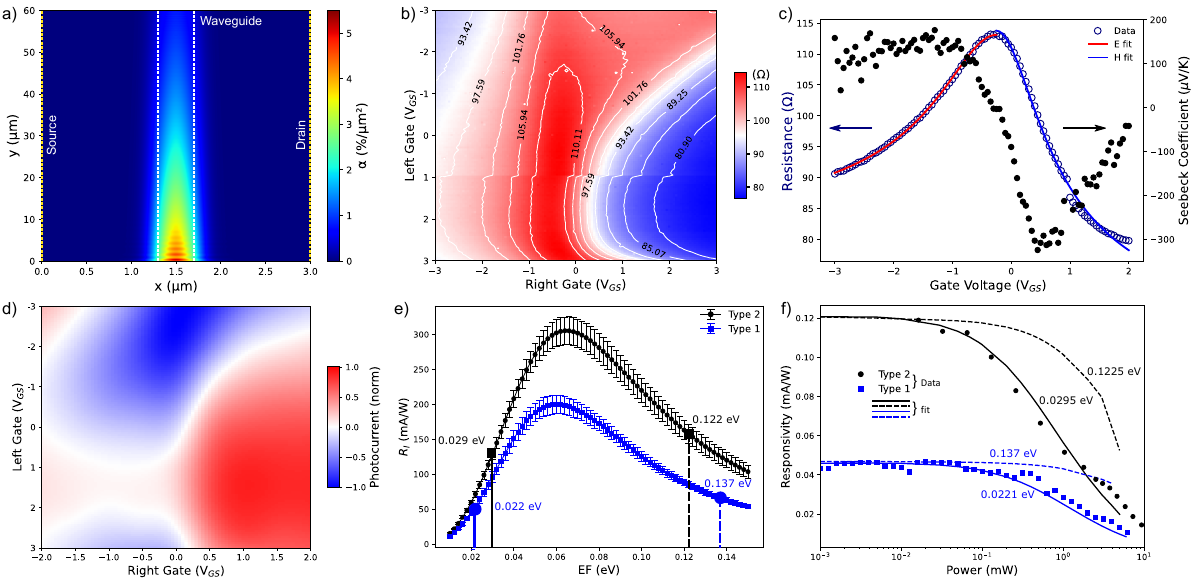}
    \caption{ Electrical and optoelectronic characterization of graphene heterostructure PDs. \textbf{a} Optical (FDTD) simulation of the photodetector, illustrating the device geometry and absorption-density profile in graphene. \textbf{b} Resistance map obtained by sweeping the left and right gate voltages (V$_\mathrm{LG}$ and V$_\mathrm{RG}$); the characteristic p–n pattern peaks at $\sim112~\Omega$. \textbf{c} Resistance as a function of gate voltage for V$_\mathrm{LG}$ = V$_\mathrm{RG}$, along with the extracted Seebeck coefficient (\( S \)). \textbf{d} Photocurrent map at zero bias as a function of V$_\mathrm{LG}$ and V$_\mathrm{RG}$, showing the characteristic sixfold pattern of PTE photodetection and locating the operating gate voltage (OGV). \textbf{e} Simulated responsivity vs.\ $E_F$ along the antisymmetric-gating configurations ($V_{\mathrm{LG}} = -V_{\mathrm{RG}}$) for Type~1 (blue) and Type~2 (black). Both peak at $E_F \approx 0.07~\text{eV}$. Solid vertical lines: operating $E_F$ (Type~1: $0.022~\text{eV}$; Type~2: $0.029~\text{eV}$) on the lower-$E_F$ branch. Dashed vertical lines: matched-$R_I$ points on the high-$E_F$ branch. Error bars: statistical variation under $\pm 10{,}000~\text{cm}^2\text{V}^{-1}\text{s}^{-1}$ and $\pm 1~\text{ps}$ operation-range spreads of $\mu(E_F)$ and $\tau_c(E_F)$. \textbf{f} Power dependence at the OGV for Type~1 and Type~2. Solid: simulation at the operating $E_F$ (lower-$E_F$ branch). Dashed: simulation at the matched-$R_I$ high-$E_F$ branch --- same low-power responsivity, distinguishable saturation behaviour. Data follow the solid curves, confirming operation at $E_F \approx 0.022~\text{eV}$ (Type~1) and $0.029~\text{eV}$ (Type~2).}
    \label{fig:pd}
\end{figure}

Prior to device fabrication, the substrate and capping-layer thicknesses and the device dimensions were optimised by COMSOL simulations to maximise the optical absorption in graphene for both Type~1 and Type~2 (see Supp.~Note~1). The source--drain distance was fixed at $W = 3~\mu\mathrm{m}$ and the channel length along the waveguide at $L = 60~\mu\mathrm{m}$, balancing channel resistance (which decreases with longer $L$) and the diminishing returns of the graphene absorption tail at large $L$. The silicon waveguide width is $w = 450~\mathrm{nm}$, corresponding to a Gaussian optical-mode FWHM of $\sim 0.42~\mu\mathrm{m}$ at the graphene level ($\sigma \approx 180~\mathrm{nm}$). Fig. \ref{fig:pd}a shows the simulated TE-mode absorption profile yielding an overall $\alpha_{\mathrm{SLG}} \approx 60\%$; the dotted rectangle outlines the waveguide/channel area.

 We evaluated the fabricated split gates by measuring under $5~\text{mV}$ source-drain bias as a function of the two gate voltage V$_\mathrm{LG}$, V$_\mathrm{RG}$. The contour map in Fig.~\ref{fig:pd}b shows the characteristic p-n pattern with a peak resistance of $\sim112~\Omega$ at the charge-neutrality. From the resistance R(V$_\mathrm{G}$) along the symmetric line V$_\mathrm{LG}$ = V$_\mathrm{RG}$ (Fig.~\ref{fig:pd}c), we extracted two-terminal hole and electron mobilities using a model\cite{Miseikis2020_UltrafastZeroBiasPD} as $\sim37{,}000~\text{cm}^2\text{V}^{-1}\text{s}^{-1}$ and $\sim43{,}000~\text{cm}^2\text{V}^{-1}\text{s}^{-1}$, respectively, with an $n^* \sim 2.1 \times 10^{11}~\text{cm}^{-2}$ for the Type 2 PD - a lower bound on the channel mobility, owing to the contact-resistance contribution embedded in the two-terminal readout. The shape of $R(V_G)$ yields, via the Mott formula\cite{Gabor2014_HotCarrierGraphene}, a Seebeck coefficient $S \approx 300~\mu\mathrm{V/K}$, consistent with Seebeck saturation at the high mobilities of our devices. Decomposing the two-terminal device resistance into channel and contact contributions yields the channel resistance $R_G \approx 15~\Omega$ and the contact resistance $R_C \approx 75~\Omega$ total at the OGV (ratio $R_C/R_G \approx 5$). The device therefore operates in the contact-resistance-dominated regime in which the $R_I^{\infty} \propto \sqrt{\tau_c/\mu}$ scaling derived in the cooling-length section applies.

 The photocurrent map acquired at zero bias by scanning both gate voltages (Fig. \ref{fig:pd}d) exhibits the characteristic sixfold pattern of PTE photodetection --- six alternating-sign lobes in the $(V_{LG}, V_{RG})$ plane --- that identifies PTE as the dominant photogeneration mechanism and locates the OGV. Numerical thermoelectric simulations using the measured $(\mu, \tau_c, n^*)$ of each heterostructure as inputs (see Methods, Ref.\cite{Ioannis_2022}) predict peak responsivities of $\sim 0.2~\mathrm{A/W}$ (Type~1) and $\sim 0.3~\mathrm{A/W}$ (Type~2) at $E_F \approx 0.07~\mathrm{eV}$ along the antisymmetric-gating configurations $V_{LG} = -V_{RG}$ (Fig. \ref{fig:pd}e); error bars reflect the standard deviation under the $\pm 10{,}000~\mathrm{cm}^2\mathrm{V}^{-1}\mathrm{s}^{-1}$ and $\pm 1~\mathrm{ps}$ measurement uncertainties on $\mu$ and $\tau_c$, respectively. These peak values are at the model-optimum $E_F \approx 0.07~\mathrm{eV}$; the OGV operating point is identified separately below.

Power-dependence measurements at zero bias (Fig. \ref{fig:pd}f) yield low-power responsivities of $\sim 0.045~\mathrm{A/W}$ (Type~1) and $\sim 0.12~\mathrm{A/W}$ (Type~2), after accounting for input-path optical losses (see Methods). Two gate-voltage configurations consistent with these low-power values exist for each device: one on the lower-$E_F$ side and one on the higher-$E_F$ side of the simulated $R_I(E_F)$ peak (Fig. \ref{fig:pd}e). The two candidates have distinguishable saturation behaviour at higher power --- a consequence of the Fermi--Dirac broadening of $\sigma(E_F, T_e)$ (see Methods). The measured power dependence follows the lower-$E_F$ branch (Fig. \ref{fig:pd}f, solid lines), identifying the OGV at $E_F \approx 0.022~\mathrm{eV}$ (Type~1) and $E_F \approx 0.029~\mathrm{eV}$ (Type~2). Crucially, Type~2 exceeds Type~1 in measured responsivity despite its (relatively) lower mobility --- consistent with the ordering predicted by the $R_I^{\infty} \propto \sqrt{\tau_c/\mu}$ scaling derived above, with magnitudes reproduced by the full numerical model at the OGV (Fig. \ref{fig:pd}f). The C-band responsivity of the Type~2 device exceeds that of previously reported PTE-based top-illuminated or planar-waveguide graphene PDs\cite{Shiue2015_GrapheneBNPD,Muench2019_WaveguidePlasmonicGraphenePD,Miseikis2020_UltrafastZeroBiasPD,Yu2023_ZeroBiasPNPD,Ma2019_PlasmonicEnhancedGraphenePD,Wang2016_SlotGraphenePD,Schuler2021_MicroringPD,Marconi2021_PAM4GraphenePD,koepfli_controlling_2024}, and is comparable to ring-resonator-based PDs (which inherently sacrifice bandwidth\cite{Schuler2021_MicroringPD}). The Johnson-limited noise-equivalent power (NEP) ranges from $\sim 3.4 \times 10^{-10}~\mathrm{W\,Hz^{-1/2}}$ at the OGV to $\sim 1 \times 10^{-10}~\mathrm{W\,Hz^{-1/2}}$ in the linear power range.

\section*{High-speed photoresponse and data detection}

\begin{figure}[h!]
    \centering
    \includegraphics[width=1\linewidth]{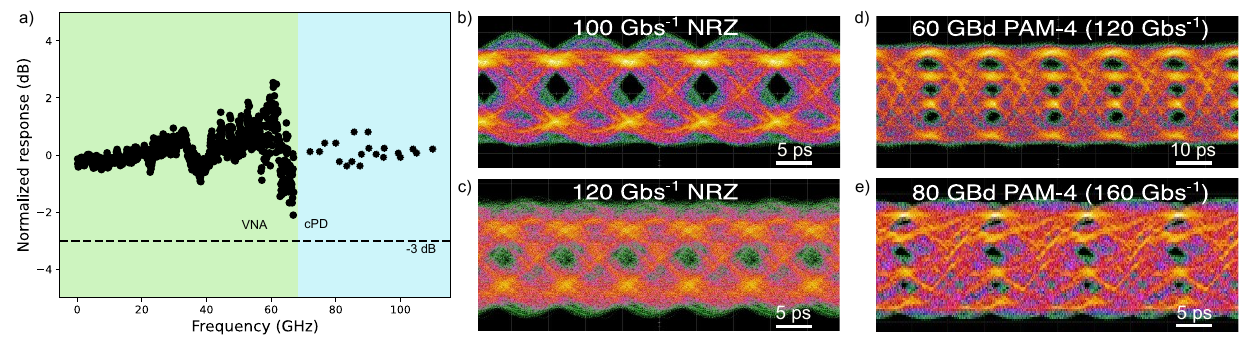}
    \caption{Frequency response of the PD measured with two complementary techniques: VNA (green-shaded region, up to $65~\text{GHz}$, set-up-limited) and dual-wavelength heterodyne (blue-shaded region, $75$--$110~\text{GHz}$). The black dotted line indicates the $-3~\text{dB}$ guideline. The Type~2 PD shows no detectable roll-off up to $110~\text{GHz}$. \textbf{b} and \textbf{c} Measured eye diagrams at 100 and 120$~\text{Gbs}^{-1}$ using NRZ modulation. \textbf{d} and \textbf{e} Eye diagrams measured using PAM-4 at 60 and 80~GBd ($120$ and $160~\text{Gbs}^{-1}$, respectively).}
    \label{fig:band}
\end{figure}

Next, we investigated the optoelectronic bandwidth of the Type~2 PD to assess the speed of the fabricated waveguide-integrated device. Two distinct setups (See Supp. Note 4) were used. First, a vector-network analyzer (VNA) was employed, with one port connected to a lithium–niobate Mach–Zehnder modulator (MZM) to modulate a CW laser source up to $67~\text{GHz}$. The output of the MZM was coupled to the device via vertical coupling between a single-mode fiber (SMF) and an on-chip grating coupler (GC). The data points in the green-shaded region of Fig.~\ref{fig:band}a represent the measured frequency response of the PD (denoted VNA). The data exhibit a flat response up to $65~\text{GHz}$ with no measurable roll-off. The observed signal drop at $> 65~\text{GHz}$ is attributed to the bandwidth limitation of the VNA. To measure beyond this range, we used an optical heterodyne setup consisting of a dual-wavelength (DW) frequency-locked laser source with tunable frequency spacing. This source was generated using a lithium–niobate phase modulator (PM) with a $40~\text{GHz}$ bandwidth, driven by a tunable frequency synthesizer. An optical programmable filter was then used to select two comb lines separated by the beat frequency $f_{\mathrm{beating}}$\cite{Montanaro2023_SubTHZWirelessGrapheneMixer}. We swept $f_{\mathrm{beating}}$ in the W-band ($75~\text{GHz}$ to $110~\text{GHz}$) with $5~\text{GHz}$ steps\cite{Marconi2021_PAM4GraphenePD}. Prior to the measurement, a commercial fast PD (Finisar XPDV4120R, denoted cPD) was used to calibrate the setup (See Supp. Note 5). The blue-shaded region in Fig.~\ref{fig:band}a illustrates the frequency response of the graphene PD up to $110~\text{GHz}$ after calibration and shows no roll-off within the measured range (setup-limited). Although contact resistance contributes to the frequency response, the absence of roll-off up to 110 GHz indicates that hot-carrier dynamics do not limit the bandwidth.

Subsequently, data detection was performed using two distinct approaches. A $256~\text{GS/s}$ arbitrary waveform generator (AWG) (Keysight M8199B) was used to encode a pseudorandom binary sequence (PRBS) into non-return-to-zero on–off keying (NRZ OOK) and four-level pulse-amplitude modulation (PAM-4). The generated electrical signal was used to drive a lithium–niobate amplitude modulator with a $40~\text{GHz}$ bandwidth, which in turn modulated a continuous-wave (CW) laser source. The signal was then coupled to the PD, and the acquired sequence was subsequently amplified and processed using standard filters such as a continuous-time linear equalizer (CTLE) and a feed-forward equalizer (FFE - 9 taps), both of which were implemented within the oscilloscope software. Fig.~\ref{fig:band}b and Fig.~\ref{fig:band}c show the eye diagrams obtained using the NRZ OOK technique, displaying a clear eye opening at $100~\text{Gbs$^{-1}$}$ and $120~\text{Gbs$^{-1}$}$. From these diagrams, the signal-to-noise ratio (SNR) and bit error rate (BER) were extracted, at $100~\text{Gbs$^{-1}$}$: SNR $\sim4.37$, BER $\sim6.17 \times 10^{-6}$; at $120~\text{Gbs$^{-1}$}$: SNR $\sim2.62~\text{dB}$, BER $\sim4.35 \times 10^{-3}$, respectively. 

Eye closure at high bit rates is limited by the modulator and RF probe bandwidth. Fig.~\ref{fig:band}d and Fig.~\ref{fig:band}e present eye diagrams for the PAM-4 modulation at $60$ and $80~\text{GBaud}$, respectively, corresponding to a bit rate of $160~\text{Gbs$^{-1}$}$. With a linear filter, we obtain a BER of $\sim2.3 \times 10^{-3}$ and $\sim6.7 \times 10^{-3}$, respectively. These values represent a lower bound owing to the setup bandwidth limitation but it is still below the soft decision FEC (SD-FEC) threshold. To the best of our knowledge, this is the highest bit rate achieved by any graphene-based direct receiver at C-band. On the other hand, Type~1 exhibited a data rate of up to $64~\text{Gbs$^{-1}$}$ and $128~\text{Gbs$^{-1}$}$ using NRZ OOK and PAM-4 modulation (See Supp. Note 5), with corresponding BER of $2.93 \times 10^{-3}$ and $4.06 \times 10^{-2}$ as the eye diagram began to close, mainly limited by the setup bandwidth and responsivity. The device also supports higher-order modulation: PAM-8 eye diagrams at 16 and 32~GBaud (48 and 96~Gbs$^{-1}$) are presented in Supp. Fig.~13d,e, demonstrating compatibility with advanced modulation formats relevant to future 200~Gbs$^{-1}$-per-lane standards (IEEE~802.3dj). Notably, at 100~Gbs$^{-1}$ NRZ, the BER of \(6.17 \times 10^{-6}\) is well below the hard-decision forward error correction (HD-FEC) threshold of \(3.8 \times 10^{-3}\) specified by IEEE~802.3 for 100GBASE-LR4 and 100G/400G Ethernet (Supp. Fig.~13b), indicating substantial margin at this data rate. These results place the Type~2 device in a distinctive operating regime: it converts \(\sim 10~\mu\text{W}\) of received optical power into a 160~Gbs$^{-1}$ data stream at an optical energy cost of \(\sim 0.06~\text{fJ/bit}\), with zero energy consumed at the detector and using only linear equalization---without recourse to complex digital signal processing.

\section*{Discussion}

\begin{figure}[h!]
    \centering
    \includegraphics[width=1\linewidth]{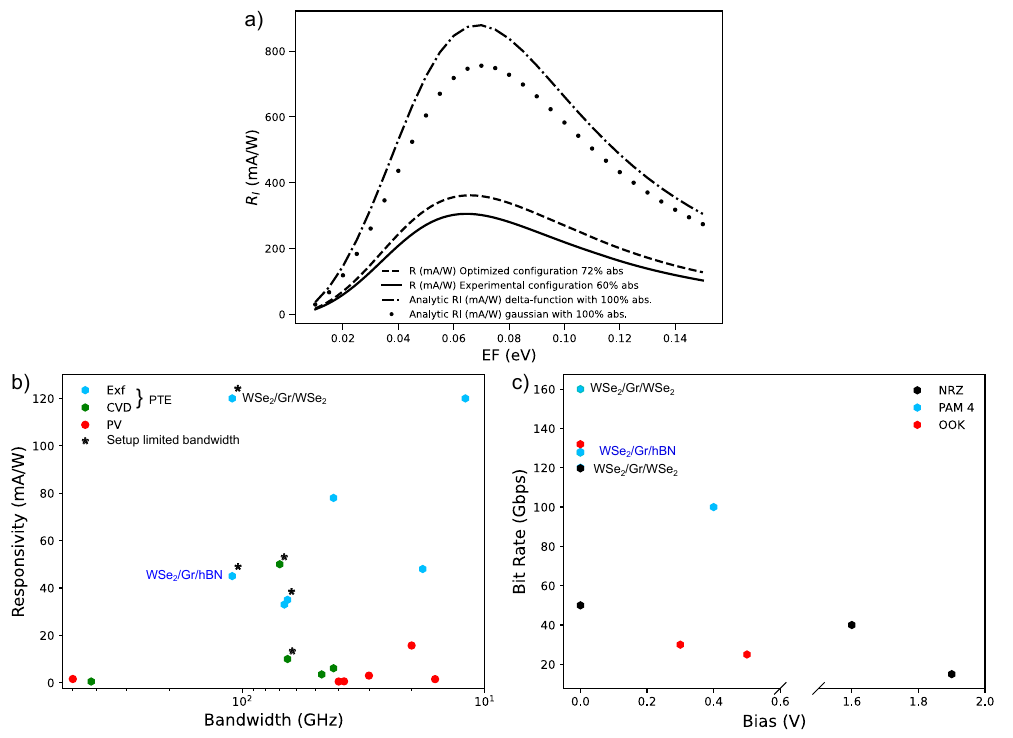}
    \caption{ Optimisation and benchmarking of Type~2 photodetectors. \textbf{a} Simulated $R_I(E_F)$ for Type~2, evaluated at the measured $(\mu, \tau_c, n^*)$; all curves peak at $E_F \approx 0.07~\text{eV}$. Solid: as-fabricated ($\sim 60\%$ absorption, numerical). Dashed: optimised ($\sim 72\%$ absorption, numerical). Dash-dot: analytical $\delta$ bound ($100\%$ absorption, Methods). Dotted: analytical Gaussian bound ($100\%$ absorption, this work, Methods). \textbf{b} Comparison of responsivity vs bandwidth for reported zero-bias graphene PDs operating in the C band. Asterisks "*"next to the dataset represent the setup-limited bandwidth. \textbf{c} Comparison of the maximum experimentally demonstrated data rates reported for waveguide-integrated graphene-based direct optical receivers. The Type~2 device presented in this work achieves one of the highest reported data rates among graphene-based direct photodetectors.}
    \label{fig:bench}
\end{figure}

To identify pathways toward the upper limit of the Type~2 PD's responsivity, we performed simulations under varying conditions using parameters extracted from our heterostructure, with the experimental configuration (black solid line in Fig.~\ref{fig:bench}a) serving as the reference. Optimizing the p–n junction width and the scalable thickness of the dielectric layer and encapsulant (See Supp. Note 6) increases absorption by 72$\%$, yielding a responsivity of $\sim$0.4 A/W. Overall, Fig.~\ref{fig:bench}a identifies potential pathways for further improving responsivity. To assess how close the simulated responsivities are to the upper bound, we benchmark them against the closed-form framework --- extending the asymmetric-contact derivation of Ref.\cite{Ioannis_2022} to handle finite-width illumination via a reduction functional $F[a]$ (see Methods):
\[
R_I = R_I^{\infty} \cdot F[a] \cdot \left(1 - e^{-L/\psi}\right)
\tag{1}
\]
where $R_I^{\infty} = gS/\beta$ is the ideal current responsivity of Ref.\cite{Ioannis_2022} in the long-channel limit ($L \gg \psi$) under $\delta$-function illumination, with $\beta = \sqrt{c_e \kappa_e/\tau_c}$ the heat-exchange coefficient introduced in Ref.\cite{Ioannis_2022}. $F[a]$ is a reduction functional accounting for the finite width and shape of the illumination profile $a(x)$, derived in this work; for the most relevant case of a Gaussian with FWHM equal to the waveguide width $w$,
\[
F[a] = F(A) = e^{A^2} \cdot \mathrm{erfc}(A), 
\qquad
A = \frac{w}{4\sqrt{\ln 2}\,\xi},
\]
with $\xi = \sqrt{\kappa_e \tau_c/c_e}$ the cooling length. The functional recovers the $\delta$-like limit $F \rightarrow 1$ for $w \ll \xi$ (maximum responsivity) and scales as $F \propto \xi/w$ for $w \gg \xi$ (diminishing response).

Evaluated with the Type~2 parameters at peak-$E_F$, the analytical current responsivity reaches $R_I^{\infty} \approx 0.9~\mathrm{A/W}$ in the $\delta$-function limit, while the Gaussian waveguide profile reduces this to $\approx 0.75~\mathrm{A/W}$ through $F(A) < 1$ --- a $\sim 16\%$ penalty that is purely geometric, arising from the ratio $w/\xi$ between the optical-mode width and the cooling length. In Fig. \ref{fig:bench}a, the two analytical bounds appear as the dash-dot ($\delta$) and dotted (Gaussian) black lines (Eq.~(1) in the $L \gg \psi$ limit, 100\% absorption); the solid and dashed curves are the numerical simulations (Methods\cite{Ioannis_2022}). All four come from the same measured $(\mu, \tau_c, n^*)$ of Type~2. Scaled to matched 72\% absorption, the analytical Gaussian bound becomes $\sim 0.54~\mathrm{A/W}$; the residual $\sim 0.14~\mathrm{A/W}$ gap to the numerical ($\sim 0.4~\mathrm{A/W}$) reflects approximations specific to the closed form --- strict low-power linearisation, step-function Seebeck profile at the junction, separation of variables, and the $W \gg \xi$ assumption that allows $T_e$ to fully relax at the source--drain contacts. The $W \gg \xi$ requirement is the structural challenge for high-$\xi$ heterostructures, most cleanly addressed by reducing $R_C$ so $W$ can scale with $\xi$ without penalizing bandwidth or current responsivity. The dominant gap from device to unmatched-absorption bound is therefore in-waveguide absorption; the hot-carrier physics established in Fig. \ref{fig:cool} is operative throughout.

The headroom identified in Fig. \ref{fig:bench}a from $\sim 0.3~\mathrm{A/W}$ (as-fabricated, $\sim 60\%$ absorption) to $\sim 0.4~\mathrm{A/W}$ (optimised, $\sim 72\%$ absorption) toward $\sim 0.75~\mathrm{A/W}$ (analytical Gaussian, 100\%) and $\sim 0.9~\mathrm{A/W}$ (analytical $\delta$, 100\%) is set primarily by in-waveguide absorption, with the residual closed by reducing the contact-resistance penalty on $W$. Throughout, the hot-carrier physics established in Fig.~\ref{fig:cool} is the enabling physics, not the limiting one. Fig.~\ref{fig:bench}b benchmarks our device against reported zero-bias graphene PDs. Prior devices combine bandwidths below $110~\mathrm{GHz}$ with responsivities below $\sim 0.08~\mathrm{A/W}$ (at OGV); our Type~2 device accesses a previously inaccessible regime at $0.12~\mathrm{A/W}$ and $>110~\mathrm{GHz}$, enabling $160~\mathrm{Gb\,s^{-1}}$ direct detection (Fig.~\ref{fig:bench}c). As mentioned earlier, both Ge-on-Si and III–V UTC photodiodes outperform our device in responsivity by \(3{-}8\times\); however, this gap is dominated by in-waveguide absorption (currently \(\sim 60\%\)) rather than by the PTE mechanism itself, and the optimization pathway in Fig.~\ref{fig:bench}a projects \(\sim 0.4~\text{A/W}\) through heterostructure design alone (Supplementary Figs.~15--17). Unlike these technologies\cite{Lischke2021_UltrafastGePD,Li2024_MUTC,Li2025_MUTC,Sun2019_HighSpeedPDs}, the PTE mechanism generates a voltage output directly via the Seebeck effect, eliminating the need of TIA. The flat S21 response to 110 GHz—with no observable RC roll-off—implies an effective device capacitance well below the geometric value, consistent with thermal localisation of the active region at high modulation frequencies. For a $160~\mathrm{Gb\,s^{-1}}$ PAM-4 link (80 Gbaud, requiring ~60 GHz system bandwidth), a co-designed voltage amplifier matched to this low capacitance is projected to consume significantly less power ($\sim$0.09 pJ/bit) than a TIA driving a reverse-biased Ge photodiode of comparable bandwidth\cite{Lischke2021_UltrafastGePD,Roelkens2024_MicroTransferPrinting}. Among other 2D materials, waveguide-integrated MoTe\(_2\)\cite{Maiti2020_MoTe2PD} and BP/MoTe\(_2\) heterojunctions\cite{Tian2022_BPMoTe2PD} offer comparable responsivities (\(0.4{-}0.7~\text{A/W}\)) but remain confined to GHz-range bandwidths.

We note that the $R_I^{\infty} \propto \sqrt{\tau_c/\mu}$ scaling derived above is specific to the contact-resistance-dominated regime ($R_C/R_G \approx 5$ at the OGV for our Type~2 PD). In a balanced regime $R_C \sim R_G$, the device conductance acquires a $\mu$-dependence and the scaling interpolates toward $\sqrt{\mu \tau_c}$ - a crossover that modulates the relative weight of $\mu$ and $\tau_c$ in the responsivity, although $\beta \propto \sqrt{\mu}$ persists in both regimes. The relative ordering of heterostructures with different $(\mu, \tau_c, n^*)$ combinations is therefore regime-dependent, and the optimum stack for a different contact regime would be selected by re-evaluating the framework with the corresponding $R_C$. The dielectric-environment handle on $\tau_c$ therefore remains operative across both regimes; what changes with $R_C/R_G$ is the quantitative weight between $\tau_c$ and $\mu$ in the responsivity hierarchy.

In summary, we demonstrate that engineering the dielectric environment of graphene---rather than maximising carrier mobility---provides an effective route to break the responsivity--bandwidth trade-off in zero-bias photothermoelectric photodetectors. WSe$_2$ encapsulation suppresses hot-carrier cooling, reducing the heat-exchange coefficient $\beta$ and yielding responsivities up to $\sim0.12~\mathrm{A/W}$ with no observable bandwidth roll-off to $110~\mathrm{GHz}$, enabling direct detection at $160~\mathrm{Gb\,s^{-1}}$ (PAM-4). These results establish dielectric control of $\beta$ as a design axis complementary to mobility and absorption, and show that device optimisation in the PTE regime requires balancing electronic quality against cooling dynamics. The compatibility of WSe$_2$ with scalable growth\cite{kim_non-epitaxial_2023,huang_wafer-scale_2026,xia_wafer-scale_2023,kwon_200-mm-wafer-scale_2024} and CMOS back-end-of-line integration\cite{yu_high-performance_2015,patoary_improvements_2023}, combined with zero-bias operation and the elimination of a transimpedance amplifier, positions this platform to address the energy and density constraints of next-generation optical interconnects.

\section*{Methods}

\subsection*{DC Measurements}

Four-probe transport measurements of the Hall-bar devices were performed by applying a low-bias (5–10 mV) between the source and drain electrodes. The resulting voltage drop was measured across the voltage probes using a lock-in. Two-probe photocurrent measurements of the sSNOM PDs were performed by applying a low-bias voltage (5–10 mV) across the device, with the resulting current change simultaneously recorded using a Keithley source-measure unit. An additional Keithley was used to apply the gate voltage and perform gate sweeps.

\subsection*{Responsivity and NEP calculation}

The optical losses in the device were quantified by defining $L_1$, $L_2$, and $L_{\mathrm{dev}}$ as the losses in the regions between the input and the PD, between the PD and the output, and within the PD itself, respectively. The total insertion loss was first measured as $10\log_{10}(P_{\mathrm{out}}/P_{\mathrm{in}})$ by coupling light into the input and collecting it at the output. This total loss can be expressed as the sum of the individual contributions, $L_1 + L_{\mathrm{dev}} + L_2$. To verify symmetry, the measurement was repeated with the propagation direction reversed (output to input), yielding identical total loss and confirming that $L_1 = L_2$. Under this condition, the total loss reduces to $2L_1 + L_{\mathrm{dev}}$, allowing $L_1$ to be extracted as $(\text{Total loss} - L_{\mathrm{dev}})/2$. The optical power incident on the detector was then determined as $P_{\mathrm{in}} - L_1$, where $P_{\mathrm{in}}$ corresponds to the optical power at the fiber output after accounting for insertion and coupling losses. The responsivity was then calculated as $R = I_{\mathrm{ph}} / (P_{\mathrm{in}} - L_1)$, where $I_{\mathrm{ph}}$ is the measured photoresponse and $P_{\mathrm{in}} - L_1$ corresponds to the optical power incident on the PD. The noise current spectral density was estimated from Johnson--Nyquist noise as 
$i_n = \sqrt{4k_B T / R}$, where $R$ is the device resistance. 
The corresponding noise-equivalent power was calculated as 
$\mathrm{NEP} = i_n / |\mathcal{R}|$, where $\mathcal{R}$ is the responsivity of the device. This expression represents the Johnson-noise-limited NEP.

\subsection*{Optical simulations}

Optical simulations were performed using the finite-difference time-domain (FDTD) method (Lumerical), with the numerical setup, mesh refinement, boundary conditions and material parameters for Au, hBN, Si and SiO$_2$ at $\lambda = 1550~\text{nm}$ identical to those described in Ref\cite{Ioannis_2022} (``Optical Modeling''). For WSe$_2$ at $\lambda = 1550~\text{nm}$ — below the optical bandgap, so absorption is negligible and we used a uniaxial dielectric with in-plane refractive index $n_{\perp} = 4.0$ and out-of-plane $n_{\parallel} = 2.656$. Graphene was implemented as a 2D conductive surface using the Kubo formalism\cite{Ioannis_2022} with $\tau_{\text{opt}} = 200~\text{fs}$, following Ref\cite{Ioannis_2022}. The fundamental TE mode of the silicon waveguide provided the excitation. The spatial absorption density $\alpha(x,y)$ in the graphene channel was integrated to yield the total absorption $\alpha_{\text{SLG}}$ and served as the optical input to the numerical thermoelectric simulations below.

\subsection*{Numerical thermoelectric simulations}
Numerical thermoelectric simulations followed the self-consistent scheme of Ref. \cite{Ioannis_2022} (``Thermoelectric Modeling''). The FDTD-computed absorption density $\alpha(x,y)$ enters as the local heat source in the steady-state heat-dissipation equation for the electronic temperature $T_e$,
\[
- \nabla \cdot (\kappa_e \nabla T_e) = - \nabla \Pi \cdot j_q - \tau_c^{-1} c_e (T_e - T_0) + \alpha(x,y) P_{\text{in}}, \tag{3}
\]
where $\kappa_e$ is the electronic thermal conductivity, $\Pi = S T_e$ is the Peltier coefficient, $j_q = -\sigma S \nabla T_e$ is the local thermoelectric current, $\sigma$ the electrical conductivity, $c_e$ the electronic specific heat, and $P_{\text{in}}$ the input optical power. The graphene transport and thermal parameters $(\sigma, S, c_e, \kappa_e)$ are evaluated self-consistently as functions of the local $E_F$ and $T_e$ using the closed-form expressions of Ref. \cite{Ioannis_2022}; the lattice temperature is held at $T_0 = 300~\text{K}$ given the much larger lattice heat capacity. The measured $\mu$, $\tau_c$, $n^*$ and contact resistance $\rho_C$ of each heterostructure serve as experimental inputs. The mobility $\mu$ and cooling time $\tau_c$ are taken as gate-voltage-operation-range averages (treated as $E_F$-independent in the simulation); their statistical spreads propagate to the responsivity error bars in Fig.~\ref{fig:pd}e. The resulting $T_e$ distribution is integrated to give the open-circuit photovoltage \[
V_{\text{PTE}} = \frac{1}{L} \iint S \nabla T_e \, dx \, dy
\] and the short-circuit photocurrent $I_{\text{PTE}} = V_{\text{PTE}} / R_D$ with $R_D = R_G + R_C$ the total device resistance. External responsivities are $R_{V,\text{ext}} = V_{\text{PTE}} / P_{\text{in}}$ and $R_{I,\text{ext}} = I_{\text{PTE}} / P_{\text{in}}$. This numerical scheme which does not assume a $\delta$-like or Gaussian absorption profile and retains the full $E_F$, $T_e$-dependence of all material parameters produces the simulated curves in Fig.~\ref{fig:pd}e and \ref{fig:pd}f (solid and dotted lines), and the ``experimental'' and ``optimised'' configurations in Fig.~\ref{fig:bench}a. The parameter-sweep maps in Figs.~\ref{fig:cool}e and \ref{fig:cool}f, and the idealised analytical bounds in Fig.~\ref{fig:bench}a, come instead from the closed-form framework described below. As $\sigma$, $S$, $c_e$, $\kappa_e$ and $\alpha$ are all recomputed self-consistently at each iteration as functions of the local $(E_F, T_e)$, the model captures the power dependence explicitly. In particular, the thermal conductivity $\kappa_e = L_0 \sigma T_e$ depends on $T_e$ through both the explicit prefactor and the Fermi--Dirac broadening of $\sigma(E_F, T_e)$: at lower $E_F$ the broadening tail spans the Dirac point, so a rising $T_e$ thermally activates additional carriers and $\sigma$, $\kappa_e$ and $\beta$ all climb rapidly with optical power, so the responsivity saturates earlier. At higher $E_F$, $k_B T_e \ll E_F$ throughout the operating range, $\sigma$ remains nearly $T_e$-independent, and the linear regime extends to higher powers. This mechanism explains why two $E_F$ values can give indistinguishable low-power responsivity (Fig.~\ref{fig:pd}e) but visibly diverge in their high-power behaviour (Fig.~\ref{fig:pd}f), providing a diagnostic of the operating Fermi level.
\subsection*{Parameter-sweep maps and bandwidth estimator}

The parameter-sweep maps of Fig. \ref{fig:cool}e and Fig. \ref{fig:cool}f are obtained directly from the closed-form framework above. Fig. \ref{fig:cool}e evaluates $R_I^{\infty} = \frac{g S}{\beta}$ over $(\mu, \tau_c, n^*)$ with $g$, $S$ and $\beta$ each carrying their full $\mu$-dependence through $\sigma$, the Mott formula, and $\kappa_e = L_0 \sigma T_e$ \cite{Ioannis_2022} so that the upper-bound responsivity of each heterostructure is read off from its measured transport parameters. Fig. \ref{fig:cool}f evaluates the bandwidth from
\[
f_c = \left(f_{c1}^{-1} + f_{c2}^{-1}\right)^{-1}, \quad
f_{c1} = \frac{1}{2\pi R_D C}, \quad
f_{c2} = \frac{1}{\tau_c},
\]
with $C = 6~\text{fF}$ from prior device characterisation of similar geometry~[35,36]; this estimator is used for comparative purposes rather than as a device-level bandwidth prediction. The hot-carrier term $f_{c2} = \frac{1}{\tau_c}$ enters as the intrinsic PTE speed limit, following the convention of Ref.\cite{Ioannis_2022} (set by hot-carrier cooling dynamics rather than by a Lorentzian $3$-dB cut-off), while the RC load enters as the standard $3$-dB cut-off $f_{c1} = \frac{1}{2\pi R_D C}$. The harmonic combination is the appropriate form for cascaded first-order responses, equivalent to time constants adding: $\frac{1}{f_c} = \tau_c + 2\pi R_D C$. Fig. \ref{fig:cool}e, \ref{fig:cool}f are therefore analytical evaluations of the framework above, distinct from the numerical thermoelectric simulations described earlier (which produce Fig. \ref{fig:pd}e, f and the ``experimental'' + ``optimised'' curves in Fig. \ref{fig:bench}a).

\section*{Authors contribution}

F.H.L.K., K.-J.T. and K.P.S. conceived the project. K.P.S. and F.H.L.K. designed the devices. K.P.S. fabricated the devices. I.V. and E.L. performed the simulations reported. S.C. assisted with the modeling. M.M, A.M performed the initial simulations. K.P.S. measured and analyzed the Hall bars. L.O., M.C. and K.P.S. measured and analyzed the sSNOM devices. A.M., V.S., S.M.K., L.K. and K.P.S. measured and analyzed the Type 1 and Type 2 photodetectors. K.W. and T.T. synthesized the hBN crystals. S.A.T. synthesized the TMD crystals. T.R. and D.V.T provided the waveguides. E.L., M.R., J.L. and F.H.L.K. supervised the work and discussed the results. K.P.S., A.M., V.S. and F.H.L.K. wrote the manuscript. All authors contributed to the manuscript revisions. A.M. and I.V. contributed equally to this work.

\clearpage
\bibliographystyle{naturemag-achim.bst}
\addcontentsline{toc}{chapter}{Bibliography}
\phantomsection
\bibliography{Waveguide_paper1}


\end{document}